\documentclass[preprint, amsmath, amssymb, preprintnumbers, superscriptaddress, floatfix, showpacs, showkeys, aps, prl]{revtex4-1}

\usepackage{graphicx}
\usepackage{float}
\usepackage{color}
\usepackage{latexsym}
\usepackage{setspace}
\usepackage{booktabs}
\usepackage{multirow}
\begin{document}

\title{Butterfly Magnetoresistance, Quasi-2D Dirac Fermi Surfaces, and a Topological Phase Transition in ZrSiS}

\author{Mazhar N. Ali}
\affiliation{IBM-Almaden Research Center, 650 Harry Road, San Jose, CA 95120, USA.}
\affiliation{Max Plank Institute for Microstructure Physics, Weinberg 2, 06120 Halle, Germany.}
\email{maz@berkeley.edu}
\author{Leslie Schoop}
\affiliation{Max Planck Institute for Solid State Research, Heisenbergstr. 1, 70569 Stuttgart, Germany.}
\author{Chirag Garg}
\affiliation{IBM-Almaden Research Center, 650 Harry Road, San Jose, CA 95120, USA.}
\affiliation{Max Plank Institute for Microstructure Physics, Weinberg 2, 06120 Halle, Germany.}
\author{Judith M. Lippmann}
\affiliation{Max Planck Institute for Solid State Research, Heisenbergstr. 1, 70569 Stuttgart, Germany.}
\author{Eric Lara}
\affiliation{Department of Chemistry, Princeton University, Princeton New Jersey 08544, USA.}
\author{Bettina Lotsch}
\affiliation{Max Planck Institute for Solid State Research, Heisenbergstr. 1, 70569 Stuttgart, Germany.}
\author{Stuart Parkin}
\affiliation{IBM-Almaden Research Center, 650 Harry Road, San Jose, CA 95120, USA.}
\affiliation{Max Plank Institute for Microstructure Physics, Weinberg 2, 06120 Halle, Germany.}

\begin{abstract}
\indent{}\textbf{Magnetoresistance (MR), the change of a material's electrical resistance in response to an applied magnetic field, is a technologically important property that has been the topic of intense study for more than a quarter century. Here we report the observation of an unusual ``butterfly'' shaped titanic angular magnetoresistance (AMR) in the non-magnetic, Dirac material, ZrSiS. The MR is large and positive, reaching nearly 1.8 x 10$^5$ percent at 9 T and 2 K at an angle of 45$^{\circ}$ between the applied current (along the \textit{a} axis) and the applied field (90$^{\circ}$ is H parallel to \textit{c}). Approaching 90$^{\circ}$, a "`dip"' is seen in the AMR which can be traced to an angle dependent deviation from the H$^2$ law. By analyzing the SdH oscillations at different angles, we find that ZrSiS has a combination of 2D and 3D Dirac pockets comprising its Fermi surface and that the anomalous transport behavior coincides with a topological phase transition whose robust signature is evident despite transport contributions from other parts of the Fermi surface. We also find that as a function of angle, the temperature dependent resistivity in high field displays a broad peak-like behavior, unlike any known Dirac/Weyl material. The combination of very high mobility carriers and multiple Fermi surfaces in ZrSiS allow for large bulk property changes to occur as a function of angle between applied fields makes it a promising platform to study the physics stemming from the coexistence of 2D and 3D Dirac electrons.} 
\end{abstract}

\maketitle
\indent{} Unconventional magnetoresistive behavior typically falls into realm of magnetic materials, displaying negative MR effects like Giant MR and Colossal MR. The MR is defined as $\frac{\rho(H)-\rho(0)}{\rho(0)}$, where $\rho(H)$ is the resistivity in an applied field H, and is usually a very small effect in non-magnetic materials following a (1+$\mu$H$^2$) law, where $\mu$ is the carrier mobility. However recently, the MR behavior of Dirac and Weyl materials have been under investigation since the discovery of titanic MR in WTe$_2$\cite{ali2014large} and linear MR as well as negative MR in Na$_3$Bi, Cd$_3$As$_2$, NbP, and others\cite{xiong2015evidence, liang2014ultrahigh, shekhar2015extremely, huang2015observation}. While these materials have shown extremely large magnitude MR due to their very high electron mobilities, their angular magnetoresistance (AMR) has been typical of 3D Fermi Surfaces; even WTe$_2$, a layered Van der Waals material, has a very 3D electronic structure as evidenced by its moderate resistive anisotropy\cite{thoutam2015temperature}.  

\indent{}ZrSiS, a tetragonal PbFCl-type compound, was recently discovered by Schoop et al\cite{schoop2015dirac} and others\cite{xu2015two, singha2016titanic} to be a Dirac material with several Dirac electron and hole pockets comprising its Fermi surface. ZrSiS is unique among the current set of Dirac/Weyl materials for several reasons: it is the first example of a ``backyard'' Dirac/Weyl material being composed of cheap, non-toxic, earth-abundant elements. It is also stable in air, water, and simple to grow crystals of. Electronically, it has a quasi-2D electronic structure (even though it has a 3D crystal structure). As shown by Schoop et al, ZrSiS's Fermi surface is entirely made up of pockets created by linearly dispersed Dirac bands resulting in a mixture of ``2D'' pancake shaped pockets (aligned with k$_{\textit{z}}$) as well as more 3D ones. ARPES also revealed an unusual quasi-2D surface state with bulk hybridization. 

\indent{}Here we report the discovery of a butterfly shaped titanic AMR effect in ZrSiS as well as an angular dependent topological phase transition, when the current is applied along the \textit{a} axis and an applied magnetic field is swept from along the \textit{c} axis to along the direction of current. The butterfly AMR can be thought of as a convolution of 2-fold and 4-fold symmetric elements whose contributions change with temperature. Beyond these two symmetries, at fields $>$3 T, higher order texturing is visible, including a ``dip'' in the MR beginning at 85$^{\circ}$ and maximizing at 90$^{\circ}$. This dip is coincident with a change in the character of the field dependent MR for angles close to 90$^{\circ}$; the MR vs H can be described in terms of a linear component added to the typical quadratic component. Starting at 85 degrees, the linear component gains weight while the quadratic contribution decreases. By analyzing the SdH oscillations at many angles, several extremal orbits were identified and found to shift dramatically with changing angle. One peak, at 243 T, shows quasi-2D behavior by shifting like a $\approx$1/cos($\theta$) law, while another peak at 23 T shows highly isotropic 3D behavior, not shifting at all. Between 85 and 95 degrees, the 243 T orbit has a $\pi$ Berry phase (extracted by fitting the resistivity oscillations to the Lifshitz-Kosevich formula), indicating a non-trivial topology. By 80 degrees however, (coincident with the dip in the AMR and change in MR versus H behavior) the Berry phase suddenly changes, implying an angle dependent topological phase transition. This type of sharp angular dependent topological phase transition has not been theoretically predicted or reported in any other Dirac/Weyl material and is likely due to the quasi-2D nature of parts of the Fermi surface in ZrSiS. It has a remarkably clear effect on the magnetoresistance, an unexpected but robust signature visible even though it is not the only contribution to transport. Also, unlike WTe$_2$\cite{ali2015correlation}, LaSb\cite{tafti2015resistivity}, NbP\cite{shekhar2015extremely}, and other high mobility semimetals, ZrSiS shows a very unusual angle dependent maximum in its temperature dependent resistivity at high field, not seen before in any Dirac/Weyl material.  

\indent{}The zero-field temperature dependent resistivity of ZrSiS is presented in Figure 1a and the inset shows the residual resistivity at 2 K. The current is applied along the \textit{a} axis of the crystal. ZrSiS has extremely low resistivity for a semimetal at room temperature with a value of 15.5$\mu$$\Omega$cm, falling to 49(4)n$\Omega$cm at 2 K, yielding a RRR of $\approx$ 300. At room temperature other high mobility semimetals (including Dirac and Weyl materials) like Cd$_3$As$_2$, LaBi, and WTe$_2$ all have resistivities at least a 1 order of magnitude greater than ZrSiS\cite{liang2014ultrahigh, ali2015correlation, shekhar2015extremely}. At low temperature, only Cd$_3$As$_2$ has a lower resistivity (21n$\Omega$cm) than ZrSiS, being comparable to high purity Bi\cite{PhysRev.91.1060}, and oxides like PdCoO$_2$\cite{hicks2012quantum}. To the best of our knowledge, ZrSiS has the lowest resistivity of any sulfide known. 

\indent{}Figure 1b shows the 9 T resistivity versus temperature of ZrSiS at various angles of the applied field with respect to the current; 90 degrees is defined as H being parallel to the \textit{c} axis and perpendicular to I. The resistivity plateaus at low temperature at an angle of 50 degrees in a fashion similar to that of several high mobility semimetals like WTe$_2$\cite{ali2015correlation} and LaSb\cite{tafti2015resistivity} at 90$^{\circ}$. However at 60$^{\circ}$, a maximum is visible, followed by a downturn in the resistivity. This behavior has not been reported before in any Dirac/Weyl material or topological insulator. The inset shows the temperature dependent Hall coefficient under various magnetic fields ($\theta$ = 90$^{\circ}$) in ZrSiS. The Hall coefficient peaks (p-type) at 100 K in a 1 T field before rapidly decreasing with decreasing temperature until plateauing by 27 K. The Hall peak moves to higher temperature with increasing magnetic field strength. Also, with stronger fields a p-n crossover occurs at 30 K for a 3 T field and 39 K for a 4 T field, and stays n-type down to 2 K. 

\indent{} The ``butterfly'' magnetoresistance is presented in Figure 2. Panel a) shows AMR at various magnetic field strengths on a polar plot, while panel b) is a conventional plot. The polar plot illustrates the 2 and 4 fold symmetries as well as the ``dip'' occurring between 85 and 95$^{\circ}$ that begins being evident by 4 T and becomes enhanced with increasing field. This ``butterfly'' pattern has been seen previously in the in-plane AMR of underdoped (antiferromagnetic) high T$_c$ superconductors like Sr$_{1-x}$La$_x$CuO$_2$\cite{jovanovic2010anisotropy} and the anisotropic magnetoresistance of rare earth manganates and other magnetic thin films \cite{zhang2010angular}. The phenomena present in ZrSiS cannot be ascribed the same origin due to the lack of magnetism in the compound. Elemental Bi, when H is rotated in plane, exhibits 6-fold symmetry in its AMR stemming from its 3 in plane ellipsoidal pockets, as well as a breakdown of symmetry in higher magnetic fields \cite{zhu2012field, collaudin2015angle}. Recently, LaBi \cite{kumar2016observation} was shown to have a 4-fold MR dependence, however the higher order texturing found in ZrSiS, visible in panel b), do not appear to be present in LaBi. The standard plot reveals the higher order texturing, such as the splitting of the peak at near 45$^{\circ}$ into two unequal maxima, in the AMR that is clearly not belonging to either the 2 or 4 fold symmetry components. 

\indent{} The AMR was measured at several different temperatures to gain insight on temperature dependance of the effect. By fitting the AMR to a simple convolution of the two symmteries; p$_1*$sin($\theta$)$^2$+p$_2*$sin(2$\theta$)$^2$+1, the relative weights of the 2-fold and 4-fold components were extracted. This fitting was done for each magnetic field strength (0.5 T through 9 T) for several different temperatures (14 K, 20 K, 25 K, 30 K, 50 K, and 70 K) (See SI). The extracted weights are plotted, in panel c), versus field strength for each temperature. At 14 K, the 2 fold component dominates over the 4-fold component until high field ($\approx$ 8.5 T), however as the temperature increases, the four fold component dies off faster than the 2-fold component, implying the cause of the 4-fold component is distinct from the cause of the 2-fold component. 

\indent{} In order to investigate the ``dip'' around 90$^{\circ}$, the MR versus field strength was measured at many different angles at 2 K. Figure 3a shows the MR for selected angles as well as the fit of the non-oscillatory background to a combination of a linear and quadratic component; MR$_{ratio} = c_1H^1 + c_2H^2$. This was used because while at 45$^{\circ}$ the MR increases in a nearly quadratic fashion (as has been seen for WTe$_2$) by 90$^{\circ}$ the MR is subquadratic. Linear MR has been seen for other Dirac materials such as Na$_3$Bi and Cd$_3$As$_2$\cite{xiong2015evidence, liang2014ultrahigh}. c$_1$ and c$_2$ are the relative weights of each component and are plotted versus angle in panel b). When H is parallel to I, the MR increases in an almost entirely linear fashion with increasing magnetic field, but very quickly (by 12$^{\circ}$) the quadratic component begins to dominate. The weight of the quadratic component increases by more than 2 orders of magnitude at 45$^{\circ}$ (from 0.13 to 14), while the linear component increases only by a factor of 8 by 90$^{\circ}$ at 90$^{\circ}$. Surprisingly, exactly in the regime where the ``dip'' became evident in the AMR, the quadratic term suddenly dips while the linear term spikes upward (between 85 and 90 degrees). In order to be sure it was not an artifact, measurements were taken beyond 90$^{\circ}$ and the expected symmetry was seen. The sudden minimization in the quadratic component is on top of an all ready 4 fold symmetry while the spike in the linear component is on top of an apparent 2 fold symmetry, again implying different origins for each contribution. 

\indent{}By extracting the SdH oscillations from the various MR vs H loops (See SI for methodology) the origin of sudden change in the linear and quadratic contributions was investigated. Figure 4a shows the FFT of the SdH oscillations taken at 90$^{\circ}$ taken at various different temperatures. F$_{\alpha}$ denotes a peak at 23 T, F$_{\beta}$ at 243 T, and F$_{\delta}$ at 130 T. Also visible are the F$_{2\alpha}$, F$_{3\alpha}$, and F$_{4\alpha}$ and the F$_{2\beta}$ harmonics. Using the Lifshitz-Kosevich temperature reduction forumla \cite{qi2015superconductivity, murakawa2013detection, luo2015electron}, the effective masses for the 3 pockets were found (inset) to be 0.11, 0.16 and 0.27, respectively. Correspondingly (SI) the areas of the Fermi surface for the 3 orbits are 2.2$*$10$^{-3}$, 2.3$*$10$^{-2}$, and 1.2$*$10$^{-2}$\AA$^{-2}$, with small K$_F$ sizes of 2.6$*$10$^{-2}$, 8.6$*$10$^{-2}$, and 6.3$*$10$^{-2}$\AA$^{-1}$, respectively. Interestingly, the Fermi velocities are quite large, at 2.8$*$10$^5$, 6.3$*$10$^5$, and 2.7$*$10$^5$m/s, respectively, comparable to Cd$_3$As$_2$, WTe$_2$, and LaBi. However, the orbit giving rise to F$_{\alpha}$ is very close in size to another similar orbit, resulting in a convoluted peak with a broadened shape. Panel b) shows how the period of the 23 T peak appears to decay with decreasing field strength and that the positions of the maxima shift slightly with increasing temperature. Both of these behaviors imply that F$_{\alpha}$ is actually comprised of two nearly superimposed peaks which interfere with each other and give the anomalous behavior.   

\indent{}Panel c) shows the FFT of the SdH oscillations at various angles. Between 60$^{\circ}$ and 50$^{\circ}$, F$_{\alpha}$ splits into 2 nearby peaks before returning to a single peak behavior by 45$^{\circ}$. The 243 T peak is sharp and can be seen shifting to higher frequencies until 55$^{\circ}$, at which point several new peaks become visible and shift in different directions with decreasing angle. The red dashed line shows F$_{\beta}$ can be followed in its shift through the 20$^{\circ}$ data. A 2D Fermi surface, when tilted with respect to an applied magnetic field, has a peak shift following a $\approx$1/cos($\theta$) law \cite{qu2010quantum, kumar2016observation}. Panel d) shows the frequency of F$_{\beta}$ as a function of angle; at larger angles, while still following the habit of a 1/cos($\theta$) dependence, the peak does not shift as much as expected for a truly 2D Fermi surface. The inset is a fit of the data to a 1/cos function with a larger period; in a perfectly 2D Fermi surface, \textit{a} would be 1 and deviation from that can be considered a measure of the "`3D"' aspect of the pocket. Since F$_{\beta}$'s shift deviates form the 1/cos law between 20 and 30$^{\circ}$, and since the shifts can be fit with \textit{a} = 0.81, the 243 T oscillation stems from a quasi$-$2D surface. The 23 T peak, on the other hand, show almost no shift for over 50 degrees, implying it comes from a 3D and highly isotropic Fermi surface. 

\indent{}Due to the singular nature of the 243 T peak near 90$^{\circ}$, it was possible to extract the Berry phase of the orbit by directly fitting the oscillations using the formula: $\rho_{xx} = \rho_0[1+A(B,T)cos2\pi(B_F/B - \delta + \gamma)]$ \cite{murakawa2013detection, liang2014ultrahigh, tayari2015two, hu2016pi, mikitik1999manifestation}. Here 1/B$_F$ is the SdH frequency and $\delta$ is the a phase shift determined by the dimensionality of the feature; being 0 for a 2D case and $-1/8$ in the 3D case (for electrons). The Berry phase is related via $\mid\gamma-\delta\mid$ = $\mid1/2-\phi_B/2\pi-\delta\mid$; values for $\mid\gamma-\delta\mid$ which imply a nontrivial $\pi$ Berry phase is 0 and 1/8 for the 2D and 3D cases respectively. Panel e) shows the extracted $\mid\gamma-\delta\mid$ for F$_{\beta}$ between 70 and 105$^{\circ}$. Between 85 and 95 degrees, $\mid\gamma-\delta\mid$ is very close to 0; a slight deviation of 0.03 is likely caused by the quasi-2D nature of the pocket causing $\delta$ to not be exactly 0, but take on an intermediate value between 0 and 1/8. As the field is tilted away from the \textit{c} axis, a sudden change in $\mid\gamma-\delta\mid$ is seen between 85 and 80 degrees. $\mid\gamma-\delta\mid$ takes on a value of 0.3; too large of a change to solely be coming from a change in $\delta$. This requires that the Berry phase also change to a value other than $\pi$, implying a transition to a trivial topology. The inset shows the deconvoluted data for the 243 T peak showcasing the sudden and visible difference in phase of the oscillations by 80 degrees. While Cd$_3$As$_2$ has a smooth change in the Berry phase of its ellipsoidal pocket with changing angle\cite{cao2015landau, zhao2015anisotropic}, this type of sharp angular dependent topological phase transition has not been reported in any other Dirac/Weyl material (nor has it been predicted to exist by theoretical analysis) and is likely due to the quasi-2D nature of parts of the Fermi surface in ZrSiS. The remarkable effect on the magnetoresistance, however, makes it noticeable in a bulk transport property of an extremely conductive semi-metal; an unexpected but robust signature that can be seen even though there are contributions from other parts of the Fermi surface. ZrSiS is a rich system for experimental study; the low temperature high-field behavior of the resistivity as well as many aspects of the AMR are still unexplained. Since the Fermi surface is made from Dirac bands, however, it is likely that the peculiar behaviors have a nontrivial origin and will benefit from careful theoretical analysis of the electronic structure, which may be key in potentially unlocking further properties of interest. 

\newpage

\textbf{Methods}

\indent{}High quality single crystals of ZrSiS were grown in a single step synthesis as described by Schoop et al \cite{schoop2015dirac}. Single crystals were grown from the mixed elements via I$_2$ vapor transport at 1100$^{\circ}$C with a 100$^{\circ}$C temperature gradient. The crystals were obtained at the cold end and then annealed in a sealed quartz tube for 1 week at 700$^{\circ}$C.

\indent{}ZrSiS crystals were structurally and chemically characterized by powder-XRD to confirm bulk purity, single crystal XRD to determine crystal growth orientation, SEM-EDX for chemical analysis, and TEM to search for a low temperature phase transition as previously described by Schoop et al \cite{schoop2015dirac}. A Quantum Design PPMS was used for transport measurements with AC transport and resistivity options. Hall measurements were taken in a 5-wire configuration while the magnetoresistance of ZrSiS samples was measured using the 4-point probe method. Due to the extremely low resistivity of the crystals, mechanical polishing was done to thin the samples down to ~70 $\mu$m in order to obtain low noise measurements. 

\indent{}The electronic structure calculations were performed in the framework of density functional theory using the \textsc{wien2k}\cite{blaha2001} code with a full-potential linearized augmented plane-wave and local orbitals [FP-LAPW + lo] basis\cite{singh2006} together with the Perdew Burke Ernzerhof (PBE) parametrization of the generalized gradient approximation (GGA) as the exchange-correlation functional. The Fermi surface was plotted with the program \textit{Xcrysden}.

\newpage
\bibliography{Lit}
\newpage

\textbf{Addendum}

\indent{}This research was supported by IBM Research as well as the MPI for Microstructure Physics in Halle and the MPI for Solid State Research in Stuttgart, Germany. The authors would like to thank Andreas Rost for useful discussions.
 
\indent{}Author Contributions: Mazhar Ali lead the research investigation and measured the transport data. Leslie Schoop and Judith Lippmann grew the samples and carried out the electronic structure calculations. Chirag Garg and Mazhar Ali conducted the data analysis. Eric Lara assisted in preparing the experiments. All authors assisted in interpreting the results, and placing them into context. Bettina Lotsch and Stuart Parkin are the principal investigators.

\indent{}Competing Interests: The authors declare that they have no competing financial interests.

\indent{}Correspondence: Correspondence and requests for materials should be addressed to Mazhar N. Ali~(email: maz@berkeley.edu).

\newpage


\begin{figure}[h]
	\includegraphics[width=0.9\textwidth]{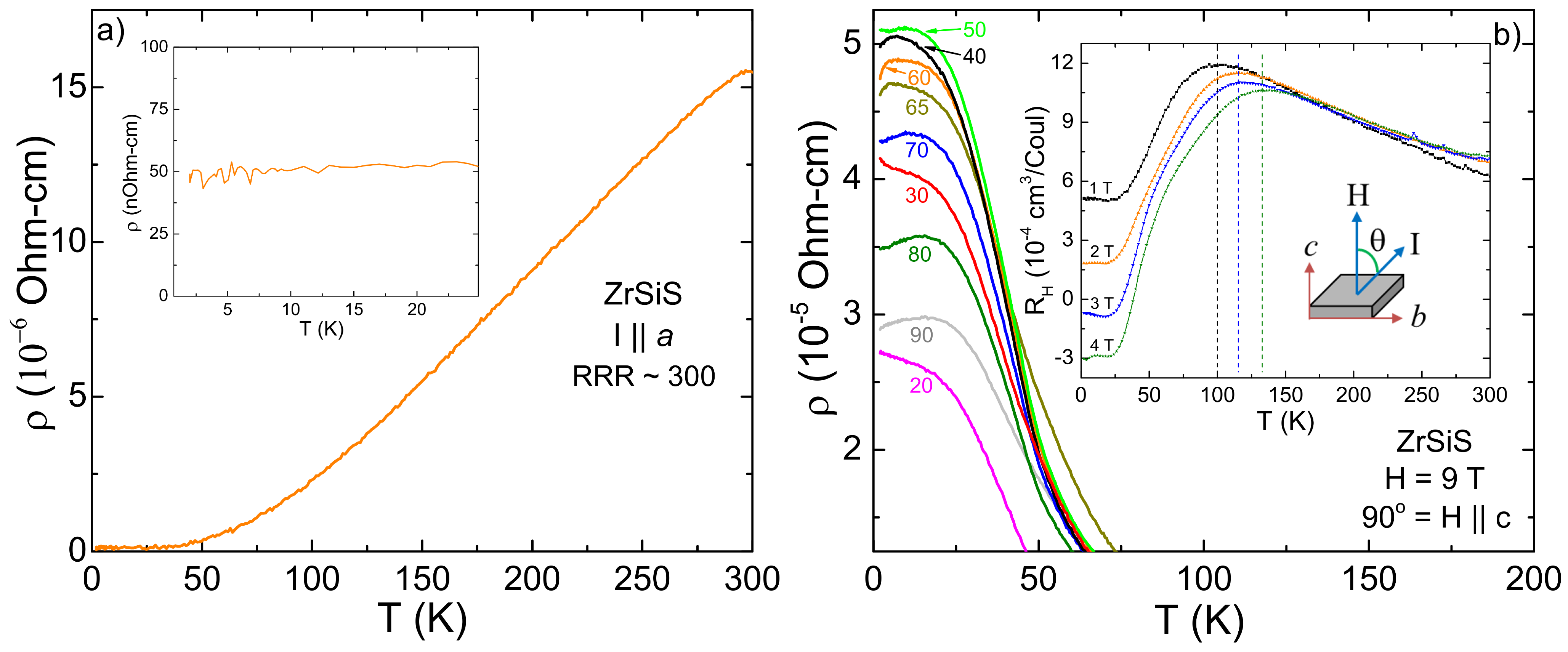}
	\caption{\scriptsize{\textbf{(color online):} Panel a) The zero field temperature dependance of the resistivity in ZrSiS. The inset shows the extraordinarily low residual resistivity of 49(4) n$\Omega$-cm for a crystal with RRR $\approx$300. Panel b) The 9 Tesla temperature dependance of the resistivity in ZrSiS at various angles. Theta is taken as the angle between the applied field and the current, which is applied along the \textit{a}$-$axis. The inset shows the Hall resistance (H perpendicular to I) as a function of temperature at various magnetic fields. Dashed vertical lines show the magnetic field dependance of the maximum R$_H$. A p-n crossover is evident for the 3 and 4 T measurements at 30 K and 39 K, respectively, before reaching a plateau at low T.}}
	\label{Figure_1}
\end{figure}
\newpage


\begin{figure}[h]
	\includegraphics[width=0.95\textwidth]{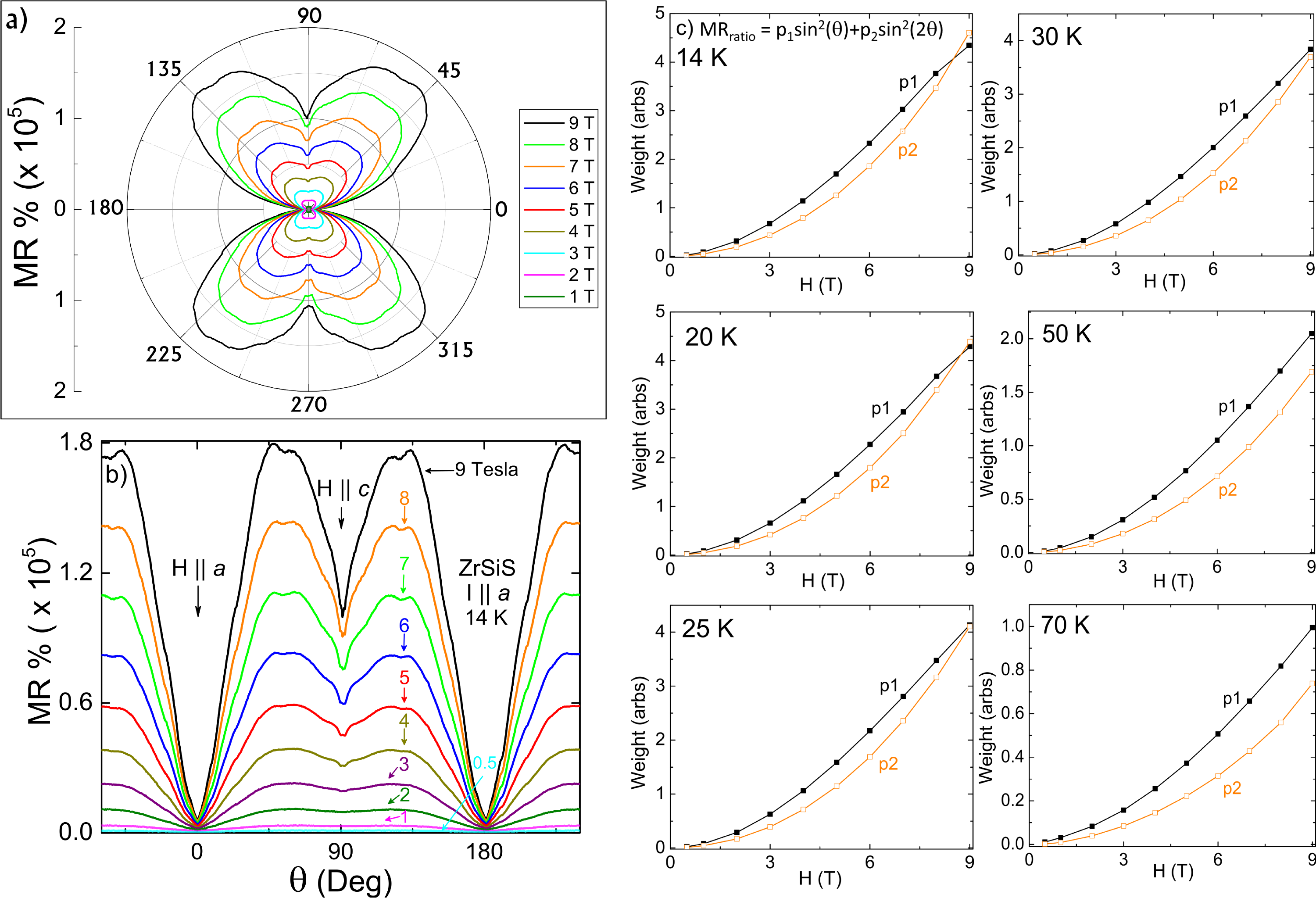}
	\caption{\scriptsize{\textbf{(color online):} The angular magnetoresistance (AMR) of ZrSiS. Panel a) Polar plot illustrating the butterfly AMR effect, coming from a convolution of 2-fold and 4-fold symmetry dependencies, taken at different applied magnetic field strengths. Theta is taken as the angle between the applied field and the current, which is applied along the \textit{a}$-$axis. Panel b) Standard plot of the AMR showcasing the field-strength dependent ``dip'' in the MR, beginning at 85$^{\circ}$, maximizing by 90$^{\circ}$, and ending by 95$^{\circ}$. Also present are additional minor oscillatory components, on top of the 2-fold and 4-fold symmetry, and peak splitting around 45$^{\circ}$ which also have a field strength dependence. Panel c) the extracted weights of the 2-fold and 4-fold symmetry components of each AMR loop between 0.5 and 9 T, at various temperatures obtained by fitting the data to a simple trigonometric equation (See SI).}}
	\label{Figure_2}
\end{figure}
\newpage


\begin{figure}[h]
	\includegraphics[width=0.95\textwidth]{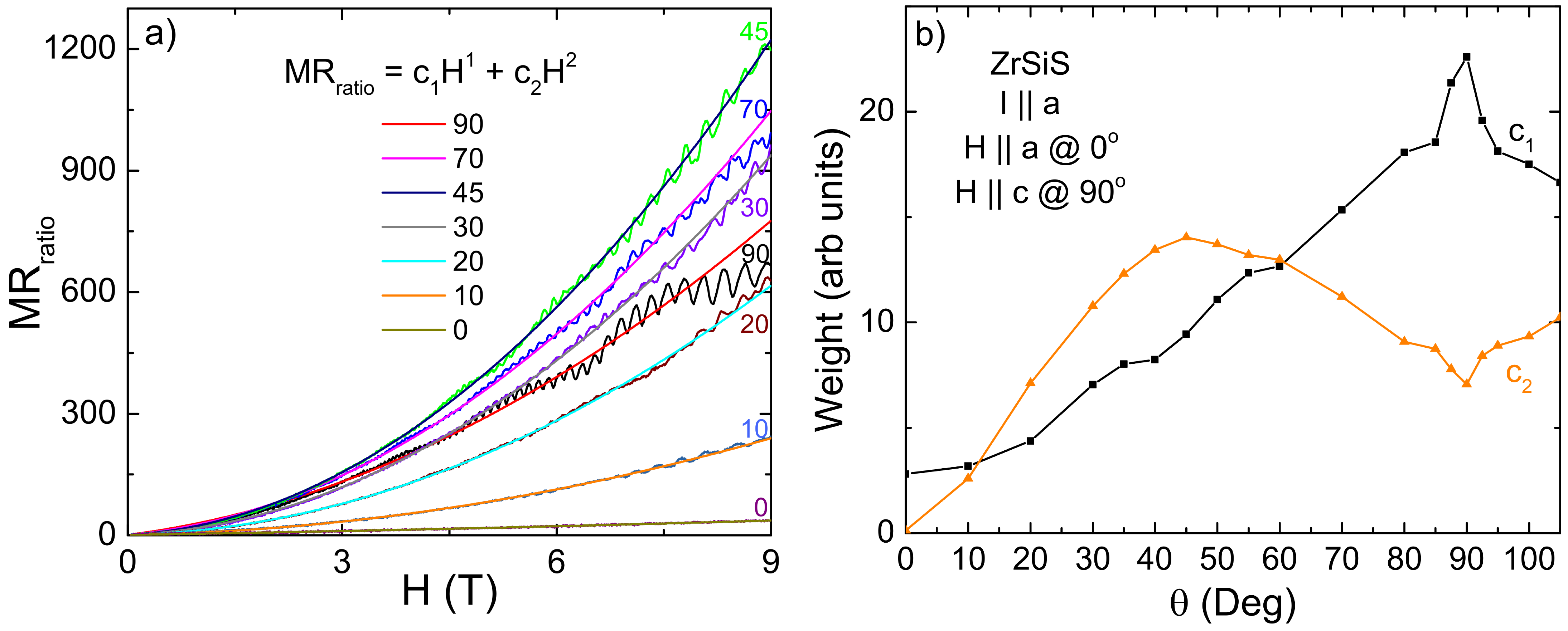}
	\caption{\scriptsize{\textbf{(color online):} The magnetoresistance as a function of applied field at various angles in ZrSiS. Panel a) the MR vs $\mu_0$H at various angles (full data in SI) along with fits to MR $=$ 1 $+$ c$_1$*H $+$ c$_2$*H$^2$, where c$_1$ and c$_2$ are the relative weights given to the linear and quadratic terms of the equation. Solid, smooth lines are the fits, while oscillating lines are data. Panel b) the extracted c$_1$ and c$_2$ dependencies from the full 0 $-$ 9T MR vs H curves taken at 0, 10, 20, 30, 35, 40, 45, 50, 55, 60, 70, 80, 85, 87.5, 90, 92.5, 95, 100, and 105 degrees where theta is taken as the angle between the applied field and the current, which is applied along the \textit{a}$-$axis. Solid squares and solid triangles are the extracted values for c$_1$ and c$_2$, respectively.}}
	\label{Figure_3}
\end{figure}
\newpage


\begin{figure}[h]
	\includegraphics[width=0.95\textwidth]{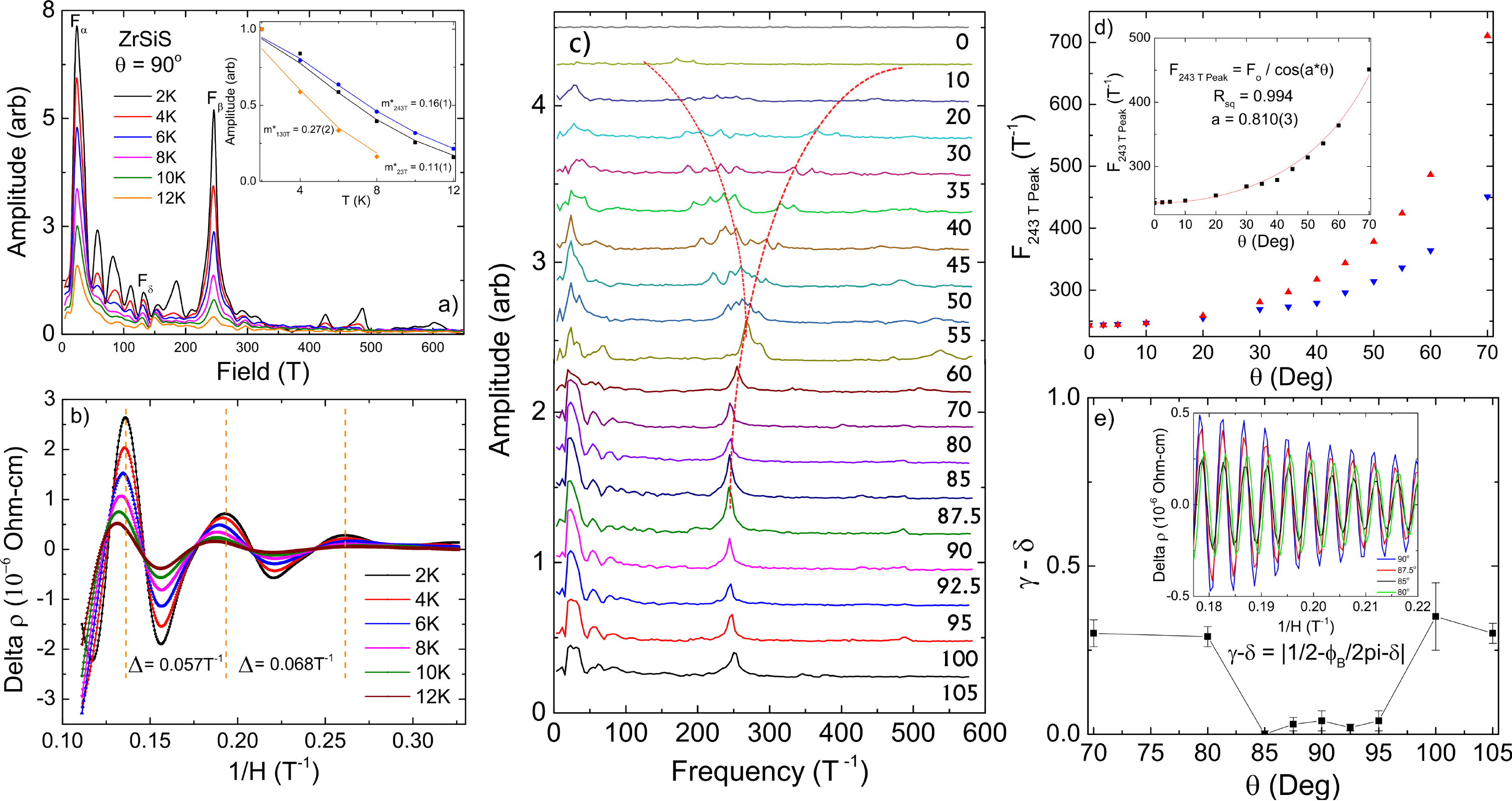}
	\caption{\scriptsize{\textbf{(color online):} The quantum oscillation analysis and extraction of the berry phase in ZrSiS. Panel a) the FFT at 90$^{\circ}$ of the SdH oscillations extracted after subtracting a 4th order polynomial background (see SI). F$_{\alpha}$, F$_{\beta}$, and F$_{\delta}$ denote frequencies at 23 T, 243 T, and 130 T, respectively. Extra peaks resulting form F$_{2\alpha}$, F$_{3\alpha}$, and F$_{4\alpha}$ as well as F$_{2\beta}$ are also visible along with 3 more unidentified frequencies at 154 T, 186 T, and 426 T. Inset, the LK fit used to determine the effective mass of the electrons for the 23 T, 243 T, and 130 T peaks. Panel b) shows the deconvoluted oscillations pertaining to the F$_{\alpha}$. Dashed lines indicate the apparent decay of the period of the oscillation with decreasing field as well as showcase the temperature dependent shift of the oscillation's maxima, indicating the presence of a nearly overlapping second oscillation which makes phase analysis of F$_{\alpha}$ extremely unreliable. Panel c) the FFT from the extracted oscillations at many different angles of the applied field. Curved lines are guides to the eye to illustrate the peak splitting and shifting occurring as a function of the angle of the applied field. Panel d) the angular dependance of F$_{\beta}$ peak. Here the definition of $\theta$ has been rotated 90$^{\circ}$ to allow for easier comparison with other materials; $\theta$ $=$ 0 is H parallel to \textit{c}. Red triangles are the expected values for a truly 2D pocket following a ~1/cos($\theta$) law, inverted blue triangles are the measured values, a deviation occurs by 20 degrees. Inset is a fit of the data to F/Bcos(a*$\theta$). Panel e) The extracted total phase ($\gamma - \delta$) of F$_{\beta}$ as a function of angle, obtained by fitting the LK formula to the deconvoluted F$_{\beta}$'s SdH oscillations. Inset is the deconvoluted data for 90, 87.5, 85 and 80 degrees, showing the sudden change in phase of the 80 degree oscillations (green line).}}
	\label{Figure_4}
\end{figure} 
\newpage

\end{document}